\PassOptionsToPackage{unicode}{hyperref}
\PassOptionsToPackage{hyphens}{url}
\PassOptionsToPackage{table}{xcolor} 

\documentclass{article}

\usepackage[hang,flushmargin]{footmisc}
\newcommand\blfootnote[1]{%
  \begingroup
  \renewcommand\thefootnote{}%
  \footnotetext{#1}%
  \addtocounter{footnote}{-1}%
  \endgroup
}

\usepackage[
    backend=biber,
    style=nature,
    sorting=nyt,
    maxbibnames=2,
    minbibnames=1,
    date=year
]{biblatex}
\AtEveryBibitem{%
  \clearfield{issn}
  \clearfield{isbn}
  \clearfield{month}
  \clearfield{day}
  \clearfield{note}
  \clearfield{editor}
  \clearlist{location}
  \clearlist{publisher}
  \clearfield{urldate}
}

\DeclareFieldFormat{note}{}
\DeclareFieldFormat{url}{}
\DeclareFieldFormat{urldate}{}
\DeclareFieldFormat{language}{}
\usepackage[a4paper, margin=2cm]{geometry}
\usepackage{setspace}
\usepackage{lineno}

\usepackage{amsmath,amssymb}
\usepackage{graphicx}
\usepackage{float}
\usepackage{xcolor}
\definecolor{deepgreen}{RGB}{0, 139, 0}
\usepackage{placeins}

\usepackage{tabularx}
\usepackage{booktabs}
\usepackage{array}
\usepackage{longtable}
\usepackage{calc}
\usepackage{etoolbox} 

\newcolumntype{C}{>{\centering\arraybackslash}X}

\makeatletter
\patchcmd\longtable{\par}{\if@noskipsec\mbox{}\fi\par}{}{}
\makeatother

\usepackage{mdframed}
\surroundwithmdframed[
  topline=false,         
  rightline=false,       
  bottomline=false,      
  leftline=true,         
  linewidth=2pt,         
  linecolor=gray!60,     
  innerleftmargin=10pt,  
  innerrightmargin=0pt,  
  innertopmargin=0pt,    
  innerbottommargin=0pt  
]{quote}

\usepackage{iftex}
\ifPDFTeX
  \usepackage[T1]{fontenc}
  \usepackage[utf8]{inputenc}
  \usepackage{textcomp}
\else
  \usepackage{unicode-math}
  \defaultfontfeatures{Scale=MatchLowercase}
  \defaultfontfeatures[\rmfamily]{Ligatures=TeX,Scale=1}
\fi
\usepackage{lmodern}

\IfFileExists{upquote.sty}{\usepackage{upquote}}{}
\IfFileExists{microtype.sty}{
  \usepackage[]{microtype}
  \UseMicrotypeSet[protrusion]{basicmath}
}{}

\IfFileExists{parskip.sty}{
  \usepackage{parskip}
}{
  \setlength{\parindent}{0pt}
  \setlength{\parskip}{6pt plus 2pt minus 1pt}
}

\makeatletter
\newsavebox\pandoc@box
\newcommand*\pandocbounded[1]{%
  \sbox\pandoc@box{#1}%
  \Gscale@div\@tempa{\textheight}{\dimexpr\ht\pandoc@box+\dp\pandoc@box\relax}%
  \Gscale@div\@tempb{\linewidth}{\wd\pandoc@box}%
  \ifdim\@tempb\p@<\@tempa\p@\let\@tempa\@tempb\fi%
  \ifdim\@tempa\p@<\p@\scalebox{\@tempa}{\usebox\pandoc@box}%
  \else\usebox{\pandoc@box}%
  \fi%
}
\def\fps@figure{htbp} 
\makeatother

\ifLuaTeX
  \usepackage{luacolor}
  \usepackage[soul]{lua-ul}
\else
  \usepackage{soul}
\fi

\usepackage{bookmark}
\IfFileExists{xurl.sty}{\usepackage{xurl}}{}
\usepackage{hyperref}
\hypersetup{
  hidelinks,
  pdfcreator={LaTeX via pandoc}
}
\IfFileExists{footnotehyper.sty}{\usepackage{footnotehyper}}{\usepackage{footnote}}
\makesavenoteenv{longtable}

\title{Beyond Point Estimates: Toward Proper Statistical Inferencing and Reporting of Intraclass Correlation Coefficients}

\author{
    Yufeng Liu\textsuperscript{1}, 
    Xiangfei Hong\textsuperscript{2,*} 
    and Shanbao Tong\textsuperscript{1,*}
}

\date{%
    \vspace{0.5em} 
    \begin{minipage}{0.9\textwidth} 
        \centering
        \small 
        \textsuperscript{1}School of Biomedical Engineering, Shanghai Jiao Tong University, Shanghai, China \\
        \vspace{3pt} 
        \textsuperscript{2}Shanghai Key Laboratory of Psychotic Disorders, Shanghai Mental Health Center, \\
        Shanghai Jiao Tong University School of Medicine, Shanghai, China 
    \end{minipage}
    }

\begin{document}

\maketitle
\blfootnote{
  \textbf{E-mail addresses}: liu.yf@sjtu.edu.cn (Yufeng Liu), 
  xfhong@sjtu.edu.cn (Xiangfei Hong), 
  stong@sjtu.edu.cn (Shanbao Tong).
}

\blfootnote{
  \textsuperscript{*}Co-corresponding authors.
}

\begin{abstract}
Reporting test-retest reliability using the intraclass correlation coefficient (ICC) has received increasing attention due to the criticisms of poor transparency and replicability in neuroimaging research, as well as many other biomedical studies. Numerous studies have thus evaluated the reliability of their findings by comparing ICCs, however, they often failed to test statistical differences between ICCs or report confidence intervals. Relying solely on point estimates may preclude valid inference about population-level differences and compromise the reliability of conclusions. To address this issue, this study systematically reviewed the use of ICC in articles published in \textit{NeuroImage} from 2022 to 2024, highlighting the prevalence of misreporting and misuse of ICCs. We further provide practical guidelines for conducting appropriate statistical inference on ICCs. For practitioners in this area, we introduce an online application for statistical testing and sample size estimation when utilizing ICCs. We recalculated confidence intervals and formally tested ICC values reported in the reviewed articles, thereby reassessing the original inferences. Our results demonstrate that exclusive reliance on point estimates could lead to unreliable or even misleading conclusions. Specifically, only two of the eleven reviewed articles provided unequivocally valid statistical inferences based on ICCs, whereas two articles failed to yield any valid inference at all, raising serious concerns about the replicability of findings in this field. These results underscore the urgent need for rigorous inferential frameworks when reporting and interpreting ICCs.

\vspace{1em}
\end{abstract}

\section{Introduction}\label{introduction}

Test-retest reliability, which assesses the stability of a measurement when repeatedly administered to the same cohorts\cite{zuo_test-retest_2014}, has gained increasing attention in light of the broader replicability crisis in science, stemming from the widespread concern that many, or even most, current published research findings are false \cite{ioannidis_why_2005}.  A \textit{Nature} survey reported that more than 70\% of researchers failed to reproduce a colleague's experiments and over half could not replicate their own \cite{baker_1500_2016}. Such limited reproducibility erodes public trust, impedes scientific progress, wastes resources, and misdirects early-career researchers \cite{pusztai_reproducibility_2013}. These concerns are particularly acute in neuroimaging research, where noisy data and conventional modeling approaches  often explain less than half of the observed variability, leaving a significant portion of this variability unexplained \cite{chen_intraclass_2018}.  The large fraction of unexplained variance not only undermines the reliability of the neuroimaging measures but also attenuates true brain-behavior associations by diminishing effect size estimates and inflating the required sample sizes \cite{gell_how_2024}.  Evaluating the reliability of measurement methods is therefore critical for ensuring that the reported findings are both robust and credible, a need that extends well beyond neuroimaging.

Early studies in psychology and neuroimaging often reported test-retest reliability using Pearson’s correlation coefficient ($r$) \cite{bruton_reliability_2000}. For example, test-retest analyses of resting-state fNIRS data initially relied on $r$ to quantify session-to-session similarity\cite{zhang_testretest_2011}, and many psychological scales historically reported reliability in terms of Pearson correlations between repeated administrations \cite{beck_inventory_1961, mohammadi_evaluation_2019}. However, Pearson’s \textit{r} only measures linear association or rank-order consistency and is insensitive to systematic differences across sessions, for example, two sets of measurements can yield $r=1.0$ even if one is a shifted or rescaled version of the other. This limitation means that \textit{r} may substantially overestimate reliability when systematic bias is present \cite{mcgraw_forming_1996}. 

The currently recommended and more widely adopted approach is to use the intraclass correlation coefficient (ICC), which is derived from variance components across repeated measurements of the same subjects \cite{muller_critical_1994, zuo_test-retest_2014}. Unlike Pearson’s $r$, the ICC simultaneously accounts for both random error and systematic sources of variance, providing a more comprehensive and interpretable index of test-retest reliability\cite{koo_guideline_2016}.

In its formal definition, ICC quantifies the proportion of total variance attributable to between-subject differences. A high ICC therefore reflects lower within-subject variability and greater temporal stability \cite{mcgraw_forming_1996}.  The framework proposed by McGraw and Wong \cite{mcgraw_forming_1996} outlines multiple ICC models tailored to different experimental designs; here we focus on the most widely applied formulation which can be expressed within an ANOVA framework \cite {zuo_test-retest_2014}. Assuming that there are $N$ subjects, each measured $k$ times, we can express the $i$-th measurement from subject $j$ by the ANOVA model: 
\begin{equation}\label{eq_intro_1}
Y_{ij}=\mu+b_{j}+w_{ij} 
\end{equation}
where $Y_{ij}$ is the observed score, $\mu$ the overall mean, $b_j$ the random effect of subject $j$ (capturing the deviation of subject $j$’s mean from $\mu$), and $w_{ij}$ the residual error of that measurement. The random effects are assumed to have variances $Var(b_j) = \sigma^2_b$ and $Var(w_{ij}) = \sigma^2_w$. Under this model, the ICC is defined as the proportion of between-subject variance to total variance:

\begin{equation}\label{eq_intro_2}
ICC = \frac{\sigma^2_{b}}{\sigma^2_{b}+\sigma^2_{w}}
\end{equation}

The ICC, when calculated from limited observations, characterizes only the sample at hand. To make valid population-level inferences, statisticians emphasize the importance of quantifying uncertainty through inferential statistics, such as confidence intervals or hypothesis tests in the frequentist framework, or credible intervals in the Bayesian framework. 

In practice, however, these inferences are frequently neglected. Many studies rely solely on point estimates of ICC to conclude population-level reliability or to compare groups and studies, without reporting uncertainty, performing formal statistical tests, or reporting confidence intervals. Such practices mask the inherent sampling variability and increase the risk of false-positive findings\cite{casella_statistical_2024, lee_alternatives_2016, cumming_new_2014, altman_why_2005}. Even when confidence intervals are reported, interpretative errors remain common, for example, the mistaken belief that overlapping confidence intervals necessarily indicate no significant difference. Misuses and misinterpretations of ICCs in these ways not only exaggerate apparent effects but also undermine the reproducibility of scientific results.

In this study, we first conducted a comprehensive review of articles published in \textit{NeuroImage}, a leading neuroimaging journal, from 2022 to 2024, focusing on the prevalence of reporting ICCs as point estimates without confidence intervals or statistical tests. Building on these findings, we advocate for the routine use of inferential approaches, like confidence intervals, significance testing and Bayesian credible intervals, when reporting and inferring ICCs. To support this practice, we developed an open-source and web-based application that enables (1) evaluating single ICCs against a reference, (2) comparing differences between two ICCs, (3) estimating confidence intervals, and (4) estimating sample size for ICC-based study designs. Using this application, we systematically re-evaluated the ICCs reported in the above \textit{NeuroImage} articles to assess whether their original claims hold under proper inferential statistical analysis. 
\section{Methods}\label{methods}

\subsection{Literature retrieval, data collection and anlysis}\label{literature-retrieval-and-search-criteria}

\begin{figure}
  \centering
  \includegraphics[width=\textwidth, height=0.6\textheight, keepaspectratio]{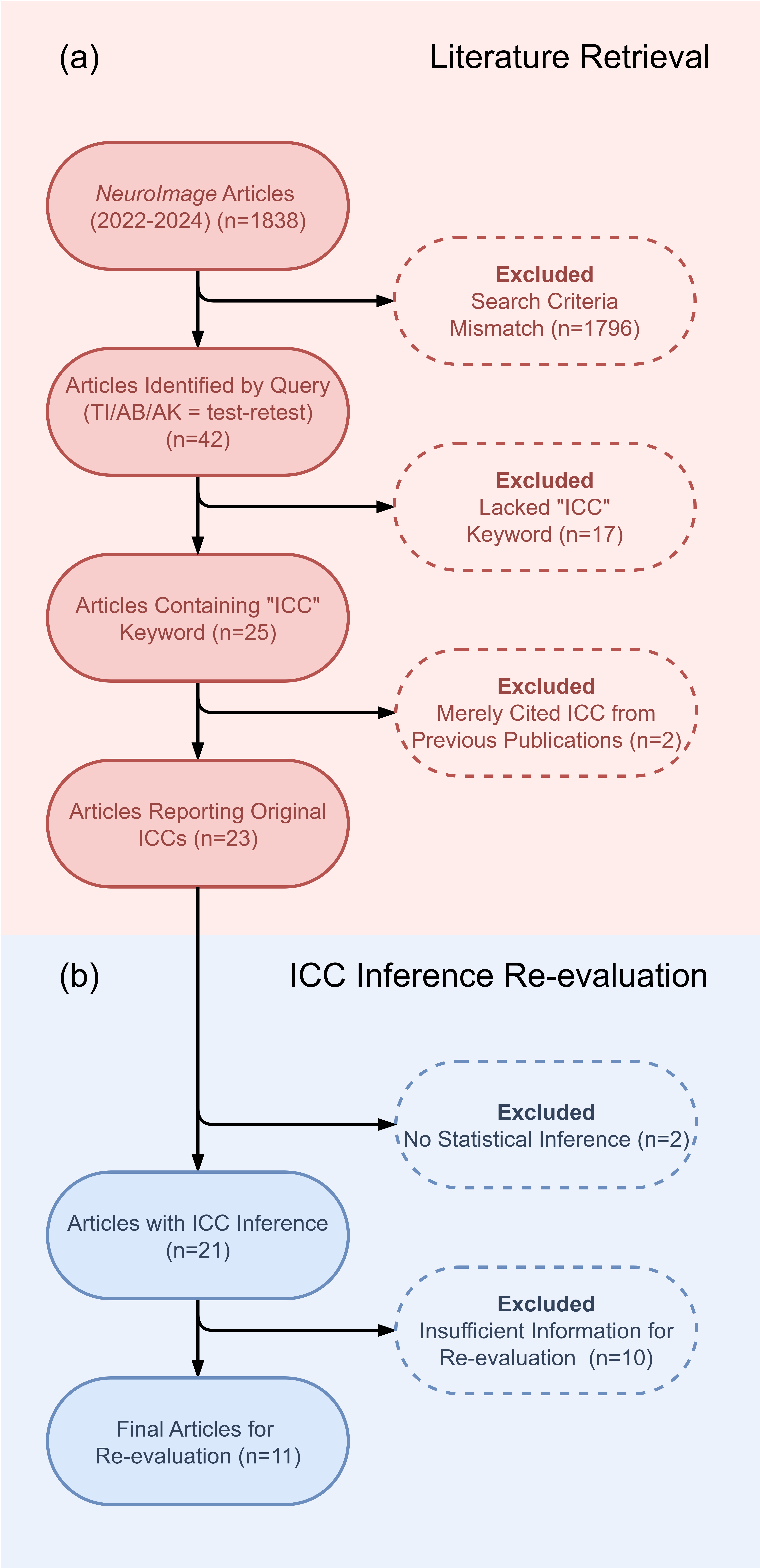}
  \caption{Flowchart of Article Retrieval and ICC Re-evaluation.
(a) Stage 1: Literature retrieval. We identified 23 \textit{NeuroImage} articles (2022–2024) that reported original ICC estimates for test–retest reliability.
(b) Stage 2: ICC inference re-evaluation. Of these 23 articles, ten provided sufficient data for re-assessing the reported ICC Inferences.}

  \label{flowchart}
\end{figure}

We systematically reviewed research articles published in \textit{NeuroImage} between 2022 and 2024. The search criteria included: \colorbox{gray!20}{
(TI=(test-retest) OR AB=(test-retest) OR AK=(test-retest)) AND
SO=(NEUROIMAGE)}. This query retrieved the papers that explicitly referenced “test-retest” in their titles, abstracts, or keywords. From these results, we retained only articles that reported original ICC calculations, excluding those that merely cited ICC values from prior publications. The article selection process is illustrated in Figure\ref{flowchart}(a).

Following article selection, we documented how ICCs were reported and applied in each study. Specifically, we recorded whether ICC values were presented as point estimates alone or accompanied by inferential statistics (e.g., confidence intervals, hypothesis tests, or Bayesian credible intervals). These ICC reports were then summarized to showcase the common gaps in the use of inferential methods.

For each reported ICC, we extracted the following information, when available:

\begin{enumerate}
\item The ICC point estimate;
\item Confidence interval(s) or Bayesian credible interval(s) for the ICC;
\item Sample size;
\item The number of retest sessions;
\item Authors’ conclusions, statements, or interpretations of the ICC values and their inferences.
\end{enumerate}

When ICCs were presented only in graphical form (e.g., bar plots), we digitized the figures using Plot Digitizer (Version 3.3.9, PRO Edition; https://plotdigitizer.com/) to obtain the corresponding numerical values. Based on the extracted information, we focused on two common types of ICC inferences: 

\begin{itemize}
\item  \textbf{Inferring test-retest reliability in population (Single ICC Evaluation)} —comparing an estimated ICC with a reference value to infer its strength (e.g., concluding that a measure is “good” when ICC > 0.75 across retests).

\item  \textbf{Inferring the difference of test-retest reliability between two populations (Two ICC Comparison)} —comparing ICCs from different measures, experimental conditions, or studies to infer which is more reliable.
\end{itemize}

In both scenarios, we reviewed whether the reported inferences relied solely on point estimates or were supported by inferential statistics, such as confidence intervals, formal null hypothesis significance testing or Bayesian credible intervals.

\subsection{OpenICC: A Web-based application for ICC inference and statistical power analysis}\label{web_based_application_for_ICC_inference}

To facilitate the reporting of reproducible ICCs in empirical studies, we developed OpenICC, an open-source, web-based application that supports statistical inference for ICCs as well as sample-size and power analysis. The application comprises four main modules:

\begin{itemize}
    \item \textbf{Single ICC Evaluation:} Calculating the confidence interval for a single ICC or performing significance testing against a reference value, to infer the magnitude of the test-retest reliability. 
    
    \item \textbf{Two Dependent ICC Comparison:}  Calculating the confidence interval for the difference between two related ICC or performing significance testing of the ICC difference.
    
    \item \textbf{Two Independent ICC Comparison:} Calculating the confidence interval for the difference between two independent ICC or performing significance testing of the ICC difference.
    
    \item \textbf{Sample Size Estimation:} Estimates the sample size required for inferences involving a single ICC and for comparisons between two ICCs, for either dependent or independent samples.
\end{itemize}

The statistical principles for each module are elaborated in the subsequent sections.Note that the sample size estimation for a single ICC was performed using the \verb|ICC.Sample.Size| R package \cite{rathbone_iccsamplesize_2015}, which implements the power analysis methods developed by Zou\cite{zou_sample_2012}. For brevity, the full formulas are not detailed here.

\subsubsection{(1) Single ICC Evaluation }\label{method-to-conduct-significance-test-of-single-icc-value}

\paragraph{Significance Testing.}
Consider a sample of $N$ participants assessed with a given instrument across $k$ repeated sessions. The goal is to determine whether the resulting test-retest reliability, quantified by the ICC, attains a satisfactory level. To address this issue, we may use the null hypothesis significance test. For the example of one-sided testing whether an ICC is greater than a reference $\rho_0$, the null and alternative hypotheses can be formulated as:

\begin{equation}
H_0: ICC \le \rho_0 \quad \text{vs.} \quad H_1: ICC > \rho_0,
\end{equation}

where $\rho_0$ is the prespecified benchmark. According to \cite{donner_testing_2002} and the analysis-of-variance (ANOVA), the sample ICC($r$) can be expressed as \eqref{eq1}:

\begin{equation}\label{eq1}
r=\frac{\mathrm{MSB}-\mathrm{MSW}}{\mathrm{MSB}+(k-1)\mathrm{MSW}},
\end{equation}

where $\mathrm{MSB}$ and $\mathrm{MSW}$ are the mean square errors, or variance, between and within subjects, respectively. To conduct the hypothesis test, a Fisher's $z$ transformation for ICCs \cite{fisher_statistical_1934,mcgraw_forming_1996,donner_testing_2002}, which differs from the standard transformation for Pearson's correlation coefficient, is used to convert $r$ to $z$ by \eqref{eqz}, which is approximately normally distributed:

\begin{equation}\label{eqz}
Z_r=\frac{1}{2}\ln \frac{1+(k-1)r}{1-r}.
\end{equation}

Under the null hypothesis, the expected mean of the sampling distribution is thus:

\begin{equation}\label{eqmean}
Z_{\rho_0}=\frac{1}{2}\ln\frac{1+(k-1)\rho_0}{1-\rho_0},
\end{equation}

and its variance is approximated by:

\begin{equation}\label{eqvar}
V_{\text{single}} =
\begin{cases}
\dfrac{k}{2(k-1)(N-2)}, & k > 2, \\
\dfrac{1}{N-\tfrac{3}{2}}, & k = 2.
\end{cases}
\end{equation}

The testing statistic $T_{Z_r}$ is then formulated as:

\begin{equation}\label{eqtz}
T_{Z_r}=\frac{Z_r-Z_{\rho_0}}{\sqrt{V_{\text{single}}}},
\end{equation}

which is approximately a standard normal distribution under $H_0$ and large $N$. If $T_{Z_r} > z_{\alpha}$, where $z_{\alpha}$ is the one-sided critical value of the standardized normal distribution at significance level $\alpha$, we reject $H_0$ and infer that the population ICC($\rho$) is greater than $\rho_0$.

\paragraph{Confidence Interval Calculation.}\label{method-to-get-the-interval-estimation-of-a-single-ICC} 
With the same Fisher's $z$ transformation for ICCs and its inverse, we can also build a confidence interval from a point estimator of ICC($r$). At the confidence level of $100(1 -\alpha)$, the interval estimation of $Z_{\rho}$ from $Z_{r}$ is:

\begin{equation}
Z_{L,U}=Z_r \pm Z_{\alpha/2}\sqrt{V_{\text{single}}}.
\end{equation}

Applying the inverse Fisher's $z$ transformation to ICCs yields the confidence interval on the original ICC scale:

\begin{equation}
\begin{aligned}
r_L =\frac{e^{2Z_L}-1}{e^{2Z_L}+k-1}, \
r_U =\frac{e^{2Z_U}-1}{e^{2Z_U}+k-1}.
\end{aligned}
\end{equation}

The Fisher's $z$ transformation for ICC is particularly suitable when only the values of $N$, $k$, and $r$ are known, and the original data are not available. When the original data are available, more robust ICC estimation and testing procedures can be implemented using R packages like \verb|irr| \cite{gamer_irr_2019}. The package operates on the ANOVA framework established by Shrout and Fleiss\cite{shrout_intraclass_1979} , where different ICC forms are calculated using the mean squares derived from specific ANOVA models (e.g., one-way or two-way).

\medskip

Note: The above the hypothesis test and the confidence interval estimation are both under the assumption of large samples. Rejecting the null hypothesis $H_0: ICC \leq \rho_0$ at level $\alpha$ is equivalent to the observation that the reference $\rho_0$ is outside of the lower bound of the $(1-2\alpha)$ confidence interval.

\subsubsection{(2) Dependent ICC Comparison }\label{method-to-test-the-equality-of-two-iccs-difference}

\paragraph{Significance Testing.}
Consider a group of $N$ participants assessed by two different instruments or under two separate conditions. The objective is to infer whether a significant difference exists between the test-retest reliability of the two resulting sets of measurements, which is commonly used for comparing the test-retest reliability of two different measures on the same cohort of subjects. Let each participant be assessed \(k_1\) times by the first instrument and \(k_2\) times by the second. We assume an equal number of measurements, i.e., \(k_1 = k_2=k\). Let \(\rho_1\) and \(\rho_2\) represent the population ICCs for the first and second instruments, respectively. The null and alternative hypotheses are formulated as:
\begin{equation}
H_0:\rho_1=\rho_2 \ vs.\ H_1:\rho_1\ne \rho_2
\end{equation}
 
Following the same procedure in a single ICC testing, we can employ Fisher's $z$ transformation (Eq. \eqref{eqz}) to the sample ICCs, $r_1$ and $r_2$ \cite{donner_testing_2002}, which results in the variables $Z_{r_1}$ and $Z_{r_2}$. 

The test statistic is based on the difference between the transformed coefficients. Under the null hypothesis $H_0 : \rho_1=\rho_2 =\rho$, the expected difference $\theta$ is: 
\begin{equation}
    \theta = E(Z_{r_1} - Z_{r_2}) = \frac{1}{2} \ln \left\{ \frac{1 + (k - 1)\rho}{1 + (k - 1)\rho} \right\}=0
\end{equation}
and the theoretical variance $V$ of this difference is:
\begin{equation}\label{dependent_v}
\mathrm{V}=\mathrm{var}(Z_{r_1}-Z_{r_2})=V_1+V_2-2\operatorname{cov}(Z_{r_1},Z_{r_2})
\end{equation}

where the theoretical covariance term is given by:

\begin{equation}\label{eqcov}
\mathrm{cov}(Z_{r_1},Z_{r_2})=\frac{k^2\rho_{12}^{2}}{2N\left\{1+(k-1)\rho_{1}\right\}\left\{1+(k-1)\rho_{2}\right\}}
\end{equation}

Here, $V_1$ and $V_2$ are the variances of $Z_{r_1}$ and $Z_{r_2}$, respectively, as defined in Eq. \eqref{eqvar}. And \(\rho_{12}\) is the population interclass correlation coefficient between measurements from the two instruments. Its sample estimate, $r_{12}$, is calculated by computing the Pearson's $r$ over a comprehensive set of paired data. Specifically, for each of the $N$ participants, every one of their $k_1$ measurements from the first instrument is paired with every one of their $k_2$ measurements from the second, which results in a total of \(Nk_1k_2\) pairs for the final calculation.  The Wald test statistic, $T_{Z_r, \text{dep}}$, is then constructed as:

\begin{equation}\label{dependent_t}
T_{Z_r, \text{dep}}=\frac{Z_{r_1}-Z_{r_2}}{\sqrt{\hat{V}}}
\end{equation}

This statistic is approximately distributed as a standard normal deviate. The term
\(\hat{V}\) is obtained by substituting
\(r_{12},r_1,r_2\) for \(\rho_{12},\rho_1,\rho_2\). The null hypothesis is rejected at a significance level of \(\alpha\) if \(|T_{Z_r, \text{dep}}|>z_{\alpha/2}\). In our verification calculations, \(r_{12}\) was set to 0 due to the unavailability of the original data's interclass correlation. Such simplification may result in more conservative findings. 

\paragraph{Confidence Interval Estimation.}\label{method-for-Confidenxe-Interval-two-iccs-difference}
Following the method proposed by \cite{ramasundarahettige_confidence_2009}, the $100(1-\alpha)\%$ confidence interval for the difference between the two ICCs ($\rho_1-\rho_2$), i.e. $[L,U]$, was calculated. Let the confidence intervals for \(\rho_1\) and \(\rho_2\) be \((l_1,u_1)\) and \((l_2,u_2)\), respectively. The lower limit, $L$, is:

\begin{equation} \label{dependent_L}
  L=r_1-r_2-\sqrt{(r_1-l_1)^2-2\widehat{\mathrm{corr}}(r_1,r_2)(r_1-l_1)(u_2-r_2)+(u_2-r_2)^2}
\end{equation}
The upper limit, $U$, is:
\begin{equation}\label{dependent_U}
U=r_1-r_2+\sqrt{(u_1-r_1)^2-2\widehat{\mathrm{corr}}(r_1,r_2)(u_1-r_1)(r_2-l_2)+(r_2-l_2)^2}
\end{equation}

where
\begin{equation}\label{dependent_corr}
\widehat{\mathrm{corr}}(r_1,r_2)=\frac{k(k-1)}{[1+(k-1)r_1][1+(k-1)r_2]}r_{12}^2
\end{equation}

\subsubsection{(3) Independent ICC Comparison}
\paragraph{Significance Testing.}
Consider a scenario with two distinct cohorts of $N_1$ and $N_2$ participants, respectively. Let participants in the first group be assessed $k_1$ times, and those in the second group be assessed $k_2$ times.  We assume an equal number of participants and measurements for both groups, i.e., $N_1=N_2=N$ and $k_1=k_2=k$. The objective is to compare the population ICCs ($\rho_1$ and $\rho_2$) from the sample ICCs ($r_1$ and $r_2$). In such a case, the formulas from the dependent test are simplified. By definition, the population interclass correlation $\rho_{12}$ is 0, which sets the theoretical covariance (eq. \eqref{eqcov}) to 0. Consequently, the test statistic (eq. \eqref{dependent_t}) simplifies to:

\begin{equation}
T_{Z_r, \text{indep}}=\frac{Z_{r_1}-Z_{r_2}}{\sqrt{\hat{V_1}+\hat{V_2}}}
\end{equation}
The resulting $T_{Z_r, \text{indep}}$ is compared to the critical value of $z_{\alpha/2}$ in  the standard normal distribution to determine the p-value.

\paragraph{Confidence Interval Estimation.}

Likewise, the $(1-\alpha)100\%$ confidence interval for the difference between the two ICCs is simplified from  (eqs.\eqref{dependent_L}, \eqref{dependent_U}) by setting the sample correlation (eq. \eqref{dependent_corr}) to 0:

\begin{equation} 
  L=r_1-r_2-\sqrt{(r_1-l_1)^2+(u_2-r_2)^2}
\end{equation}
\begin{equation}
U=r_1-r_2+\sqrt{(u_1-r_1)^2+(r_2-l_2)^2}
\end{equation}

\subsubsection{(4) Sample Size for ICC Comparisons}\label{online-tool-for-sample-size-estimation}
Because both the width of the confidence interval and the significance level (i.e., the p-value) are highly sensitive to sample size, determining an adequate sample size is a fundamental part of power analysis. The objective is to identify the minimum number of participants needed to control Type I and Type II errors for a given effect size. Accordingly, sample-size determination is a routine and essential step in experimental design.

In the OpenICC application, we thus provide the sample-size estimation procedures for comparing two ICCs. The methodology differs slightly depending on whether the ICCs are dependent (derived from the same cohort of subjects) or independent (obtained from two distinct subject groups).

\paragraph{Dependent samples.}
This scenario involves a group of $N$ participants, with each participant evaluated using two different instruments. Multiple assessments are obtained from each instrument, allowing for the calculation of an ICC for each. The sample size calculation is derived from the properties of the test statistic used to compare the two ICCs. As established in subsection  "\nameref{method-to-test-the-equality-of-two-iccs-difference}" , the analysis relies on Fisher's $z$ transformation for ICCs. Let $Z_{r_1}$ and $Z_{r_2}$ be the transformed sample ICCs, as defined above. Their difference, $(Z_{r_1} - Z_{r_2})$, follows an approximate normal distribution:
\begin{equation}
(Z_{r_1} - Z_{r_2}) \sim \mathrm{N}(\theta, V),
\end{equation}
where $\theta = E(Z_{r_1} - Z_{r_2})$ is the population mean of the difference, and $V = \mathrm{var}(Z_{r_1} - Z_{r_2})$ is its theoretical variance (defined in Eq. \eqref{dependent_v}). To achieve a desired statistical power of  $(1-\beta)$ for detecting a specified difference \(|\theta| =d\) at a significance level of $\alpha$, the following inequality must be satisfied:

\begin{equation}
d-Z_{\alpha/2}\sqrt{V} \geq Z_{\beta}\sqrt{V},
\end{equation}

where \(\alpha\) represents the probability of a Type I error, and  \(\beta\) represents the probability of a Type II error. The terms \(Z_{\alpha/2}\) and \(Z_{\beta}\) are the critical values from the standard normal distribution that correspond to upper tail probabilities of \(\alpha/2\) and \(\beta\), respectively. This inequality can be rearranged to solve for the required variance of the estimator: 
\begin{equation}
V \leq \frac{d^2}{(Z_{\beta}+Z_{\alpha/2})^2}.
\end{equation}

Assuming an equal number of measurements for both instruments $(k_1=k_2=k)$, the variance of the estimator is given by: 
\begin{equation} \label{Vvalue}
V=
\begin{cases}
    \dfrac{k}{(k-1)(N-2)}-\dfrac{k^2\rho_{12}^{2}}{N\left\{1+(k-1)\rho_{1}\right\}\left\{1+(k-1)\rho_{2}\right\}} ,& k >2,\\
    \dfrac{2}{N-\tfrac{3}{2}}-\dfrac{k^2\rho_{12}^{2}}{N\left\{1+(k-1)\rho_{1}\right\}\left\{1+(k-1)\rho_{2}\right\}},& k = 2.
\end{cases}
\end{equation}

A simplification is applied to the variance term to permit the isolation of $N$. We approximate the denominators $(N-2)$ (for $k>2$) and $(N-3/2)$ (for $k=2$) as $N$, which is asymptotically valid when the sample size is sufficiently large. This simplification unifies the distinct cases in the inequality (\ref{Vvalue}),  allowing the inequality to be expressed as:

\begin{equation}
\frac{k}{(k-1)N}-\frac{k^2\rho_{12}^{2}}{N\left\{1+(k-1)\rho_{1}\right\}\left\{1+(k-1)\rho_{2}\right\}} \leq \frac{d^2}{(Z_{\beta}+Z_{\alpha/2})^2}
\end{equation}

By rearranging the terms to isolate N, we can determine the minimum required sample size. The resulting formula is given by:

\begin{equation}\label{eqn}
N \geq (\frac{k}{(k-1)}-\frac{k^2\rho_{12}^{2}}{[1+(k-1)\rho_{1}][1+(k-1)\rho_{2}]})\cdot\frac{(Z_{\beta}+Z_{\alpha/2})^2}{d^2}
\end{equation}

\paragraph{Independent samples.}
Comparing ICCs from two independent groups follows similar principles. The key difference is that for independent samples, the interclass correlation $\rho_{12}$ is equal to 0.  By setting $\rho_{12}=0$ in Equation \eqref{eqn}, the covariance term disappears, and the formula simplifies significantly. The resulting formula gives the required sample size per group ($N$), assuming an equal sample size for both groups:

\begin{equation}\label{eqn:independent_ss}
N \geq \left(\frac{k}{k-1}\right)\cdot\frac{(Z_{\beta}+Z_{\alpha/2})^2}{d^2}
\end{equation}

\subsection{Re-evaluating the ICC inferences in \textit{NeuroImage} papers }\label{conclusion-extraction-and-validation}

We only re-evaluated the ICC inferences with complete information for verification. Specifically, the required information for recalculating the ICC confidence intervals and the corresponding null-hypothesis significance testing statistics included sample sizes, trial numbers, and specific ICC values. Figure \ref{flowchart}(b) illustrates the paper filtering process for re-evaluation. When multiple ICC-based statements were reported in a paper, we verified the most central ones (limited to a maximum of two), typically those appearing in the abstract section. If ICC was not explicitly mentioned in the abstract, we re-evaluated the key inference statements in the main text that relied on ICC. 

If an inferential statement is based on a specific ICC (or ICCs), we recalculated the confidence interval(s) and performed null hypothesis significance test(s) using OpenICC; otherwise, all relevant ICCS were verified if not specified. A complete list of the selected inferential statements and ICC values for verification is provided in Supplementary Materials (Excel file, Table S1). The hypothesis tests were tailored to the inference type: 

\begin{enumerate}
    \item Two-sided tests were used for comparing two ICCs;
    \item One-sided tests were applied when comparing a single ICC against a reference which was either cited from the original article, or by the reliability level recommendations in \cite{koo_guideline_2016}:   
    \begin{itemize}
        \item <0.5: poor 
        \item 0.5-0.75: moderate 
        \item 0.75-0.9: good
        \item >0.9: excellent
    \end{itemize}
\end{enumerate}

\section{Results}\label{result}
\subsection{Prevalence of Misusing and Misreporting of ICC in Test-Retest Studies}  
As illustrated in Figure\ref{flowchart}(a), the searching process resulted in 23 articles that reported original ICC calculations (Figure\ref{flowchart}(b)). After excluding two articles that did not involve inference, 21 articles remained that included inferential statements or analyses. Of these, only five reported confidence or credible intervals. Among the remaining 16 studies without these intervals, ten did not perform any statistical significance testing and relied solely on point estimates of the ICC for inferences. Notably, one study even misused the sample size \cite{feng_whole_2024}. 
The authors calculated ICCs for test-retest reliability by treating 1,395 regions of interest (ROIs) as independent samples. However, one hundred and fifty-five ROIs from each subject were by all means not independent at all (see Figs. 7(c) and 8(c) in \cite{feng_whole_2024}).  Four of the other six, however, conducted between-group tests, though the ICCs were derived from the same dataset \cite{aye_test-retest_2022,kennedy_reliability_2022,wang_mapping_2024,lin_voxel-based_2024}. Concerns on treating ICCs as independent observations from dependent sources in statistical testing will be discussed in the \nameref{discussion} section.

\begin{table}[htbp]
\centering
  \caption{Overview of ICC Reporting and Inference}
  \label{overview}
  
  \begin{tabularx}{0.9\textwidth}{@{}p{4cm}CCC@{}} 
    \toprule
    \textbf{Article} & \textbf{Reporting CI/CrI} & \textbf{ICC Inference} & \textbf{Inferring Method} \\
    \midrule
    Amemiya et al.,2022\cite{amemiya_reliability_2022} & \textcolor{gray}{No} & \textcolor{deepgreen}{Single/Difference} & \textcolor{gray}{Point Est.} \\
    \addlinespace 
    Aye et al.,2022\cite{aye_test-retest_2022} & \textcolor{gray}{No} & \textcolor{deepgreen}{Difference} & \textcolor{deepgreen}{Point Est./NHST$^*$}\\
    \addlinespace
    Bollack et al., 2023\cite{bollack_evaluation_2023} & \textcolor{deepgreen}{Yes} & \textcolor{deepgreen}{Single/Difference} & \textcolor{deepgreen}{CI$^*$}  \\
    \addlinespace
    Bordin et al., 2023\cite{bordin_optimal_2023} & \textcolor{gray}{No} & \textcolor{deepgreen}{Single/Difference} & \textcolor{gray}{Point Est.}  \\
    \addlinespace
    Cao et al., 2023\cite{cao_effects_2023} & \textcolor{gray}{No} & \textcolor{deepgreen}{Single/Difference} & \textcolor{deepgreen}{Point Est./NHST}\\
    \addlinespace
    De Lange et al., 2023\cite{de_lange_structural_2023} & \textcolor{gray}{No} & \textcolor{deepgreen}{Single/Difference} & \textcolor{gray}{Point Est.} \\
    \addlinespace
    Dular et al., 2024\cite{dular_base_2024} & \textcolor{gray}{No} & \textcolor{deepgreen}{Single/Difference} & \textcolor{gray}{Point Est.} \\
    \addlinespace
    Dumais et al.,2022\cite{dumais_eicab_2022} & \textcolor{gray}{No} & \textcolor{deepgreen}{Single} & \textcolor{gray}{Point Est.} \\
    \addlinespace
    Faber et al.,2022\cite{faber_cerebnet_2022} & \textcolor{deepgreen}{Yes} & \textcolor{deepgreen}{Difference} & \textcolor{gray}{Point Est.} \\
    \addlinespace
    Feng et al., 2024\cite{feng_whole_2024} & \textcolor{gray}{No}& \textcolor{deepgreen}{Single} & \textcolor{gray}{Point Est.}\\
    \addlinespace
    Flournoy et al., 2024\cite{flournoy_precision_2024} & \textcolor{deepgreen}{Yes} & \textcolor{deepgreen}{Single}  & \textcolor{deepgreen}{CrI}\\
    \addlinespace
    Hu et al.,2022\cite{hu_new_2022} & \textcolor{deepgreen}{Yes} & \textcolor{deepgreen}{Difference} & \textcolor{deepgreen}{NHST$^*$}\\
    \addlinespace
    Kai et al.,2022\cite{kai_mapping_2022} & \textcolor{gray}{No} & \textcolor{deepgreen}{Difference} & \textcolor{gray}{Point Est.}\\
    \addlinespace
    Kennedy et al.,2022\cite{kennedy_reliability_2022}& \textcolor{gray}{No} & \textcolor{deepgreen}{Single/Difference} & \textcolor{deepgreen}{Point Est./NHST$^*$}\\
    \addlinespace
    Lin et al., 2024\cite{lin_voxel-based_2024} & \textcolor{gray}{No}  & \textcolor{deepgreen}{Single/Difference} & \textcolor{deepgreen}{Point Est./NHST$^*$}\\
    \addlinespace
    Oishi et al.,2023\cite{oishi_macromolecular_2023} & \textcolor{gray}{No} & \textcolor{gray}{No} & \textcolor{gray}{/}\\
    \addlinespace
    Pirastru et al., 2023 \cite{pirastru_impact_2023}& \textcolor{gray}{No} & \textcolor{deepgreen}{Single/Difference} & \textcolor{gray}{Point Est.}\\
    \addlinespace
    Radwan et al.,2022\cite{radwan_atlas_2022} & \textcolor{deepgreen}{Yes} & \textcolor{gray}{No} & \textcolor{gray}{/}\\
    \addlinespace
    Tetereva et al.,2022\cite{tetereva_capturing_2022} & \textcolor{deepgreen}{Yes} & \textcolor{deepgreen}{Single} & \textcolor{gray}{Point Est.}\\
    \addlinespace
    Tregidgo et al., 2023\cite{tregidgo_accurate_2023} & \textcolor{gray}{No} & \textcolor{deepgreen}{Single} & \textcolor{gray}{Point Est.}\\
    \addlinespace
    Wang et al., 2024\cite{wang_mapping_2024} & \textcolor{gray}{No}  & \textcolor{deepgreen}{Single/Difference} & \textcolor{deepgreen}{Point Est./NHST$^*$}\\
    \addlinespace
    Xu et al., 2023\cite{xu_test-retest_2023} & \textcolor{deepgreen}{Yes} & \textcolor{deepgreen}{Single} & \textcolor{gray}{Point Est.}\\
    \addlinespace
    Zhao et al., 2023\cite{zhao_whole-cerebrum_2023} & \textcolor{gray}{No} & \textcolor{deepgreen}{Single/Difference} & \textcolor{gray}{Point Est.}\\
    \addlinespace
    \midrule
    \textbf{Summary} &\textbf{Yes: 31.58\%}  &\textbf{Single/Difference: 89.47\%}  &\textbf{Correct NHST/CI/CrI: 25\%} \\
    
    \bottomrule
  \end{tabularx}
    \vspace{4pt} 

  \parbox{0.9\textwidth}{
    \small 
    \textbf{Notes:} \textbf{Reporting CI/CrI} indicates whether a confidence interval or credible interval was provided. \textbf{ICC Inference} specifies whether inferential analysis on ICCs was conducted and, if so, the type: \textit{Single} = evaluation of a single ICC; \textit{Difference} = comparison of two ICCs; \textit{Single/Difference} = both types performed. \textbf{Inference Method} denotes the statistical basis for inference (\textit{Point Est.} = inference based solely on point estimates; \textit{NHST} = null-hypothesis significance testing; \textit{CI} = use of confidence intervals; \textit{CrI} = use of credible intervals). The \textit{Summary} row reports the proportion of articles that (1) reported CI/CrIs, (2) conducted any inferential analysis, and (3) used correct inferential methods among those performing inferential statistics (including \textit{NHST/CI/CrI}). An asterisk (*) in the “Inference Method” column indicates that an incorrect or flawed inferential approach was used, albeit NHST, CI or CrI was applied. 
  }
  
\end{table}

Furthermore, several forms of ICC misreporting or misuse were observed even among the five articles that reported confidence or credible intervals, including:

\begin{enumerate}
\def\labelenumi{(\arabic{enumi})}
\item
\textbf{Relying solely on point estimates for inference despite the available confidence intervals.}\cite{faber_cerebnet_2022,tetereva_capturing_2022,xu_test-retest_2023} For example, \cite{faber_cerebnet_2022} improperly drew inferences about ICCs based solely on point estimates without incorporating the available interval information. The authors stated:

\begin{quote}
    "The ICCs of $CerebNet$ and $\mathrm{ACAPULCO^{rt}}$ range between 0.635 and 0.997 across both datasets with – in most cases – more consistent results (higher ICC) for $CerebNet$. In fact, the ICC is superior for $CerebNet$ over $\mathrm{ACAPULCO^{rt}}$ in 24 of 27 sub-structures for the Kirby data and in 23 out of 27 sub-structures for the OASIS1 data set set as well as for the combined regions of the vermis and the left and right hemispheric CGM." 
\end{quote}
This approach compares methods based solely on which yields a "higher" ICC point estimate. By disregarding the confidence intervals, this reasoning fails to account for sampling variability, potentially interpreting random noise as a meaningful difference in reliability. 

\item
\textbf{Misinterpreting the meaning of overlapping confidence intervals.} One study incorrectly assumed that overlapping confidence intervals necessarily indicate a lack of statistically significant differences\cite{bollack_evaluation_2023}. Specifically, the authors stated:
\begin{quote}
    "The overlapping confidence intervals suggest that there is no evidence of difference in this across methods."
\end{quote}

This interpretation is misleading. Overlapping confidence intervals does not, in general, imply the absence of a statistically significant difference. According to \cite{cumming_inference_2005}, when comparing two independent means, statistical significance at the p $\leq$ 0.05 level is typically indicated when the overlap between the two 95\% confidence intervals is less than approximately half of the average margin of error. Although this guideline pertains to means, the conceptual principle—namely that interval overlap is not equivalent to hypothesis testing—applies similarly to ICCs.

\item
\textbf{Incorrectly implementing bootstrap procedures.} In one case, researchers used a bootstrap procedure to compare two ICCs but applied flawed inferential methods \cite{hu_new_2022}. Specifically, the authors treated the bootstrap iterations as independent observations and compared them using a two-sample t-test, stating:

\begin{quote}
    "The mean value and confidence interval of ICC(A,1) for both peak-based and pointwise analyses were obtained 1600 times by bootstrap [\dots] Two-sample t-test was applied to compare the result between peak-based and pointwise analyses."
\end{quote}

This approach is statistically invalid because the bootstrap-derived ICC values are not independent samples and should not be analyzed using a two-sample t-test. A detailed analysis of this error and the correct bootstrap procedure for comparing ICCs is provided in the "\nameref{bootstrap}" section of the Discussion.

\end{enumerate}

\subsection{The OpenICC Application}\label{R-shiny-web-application-development}

\begin{figure}[htbp]
  \centering
  \includegraphics[width=\textwidth, height=\textheight, keepaspectratio]{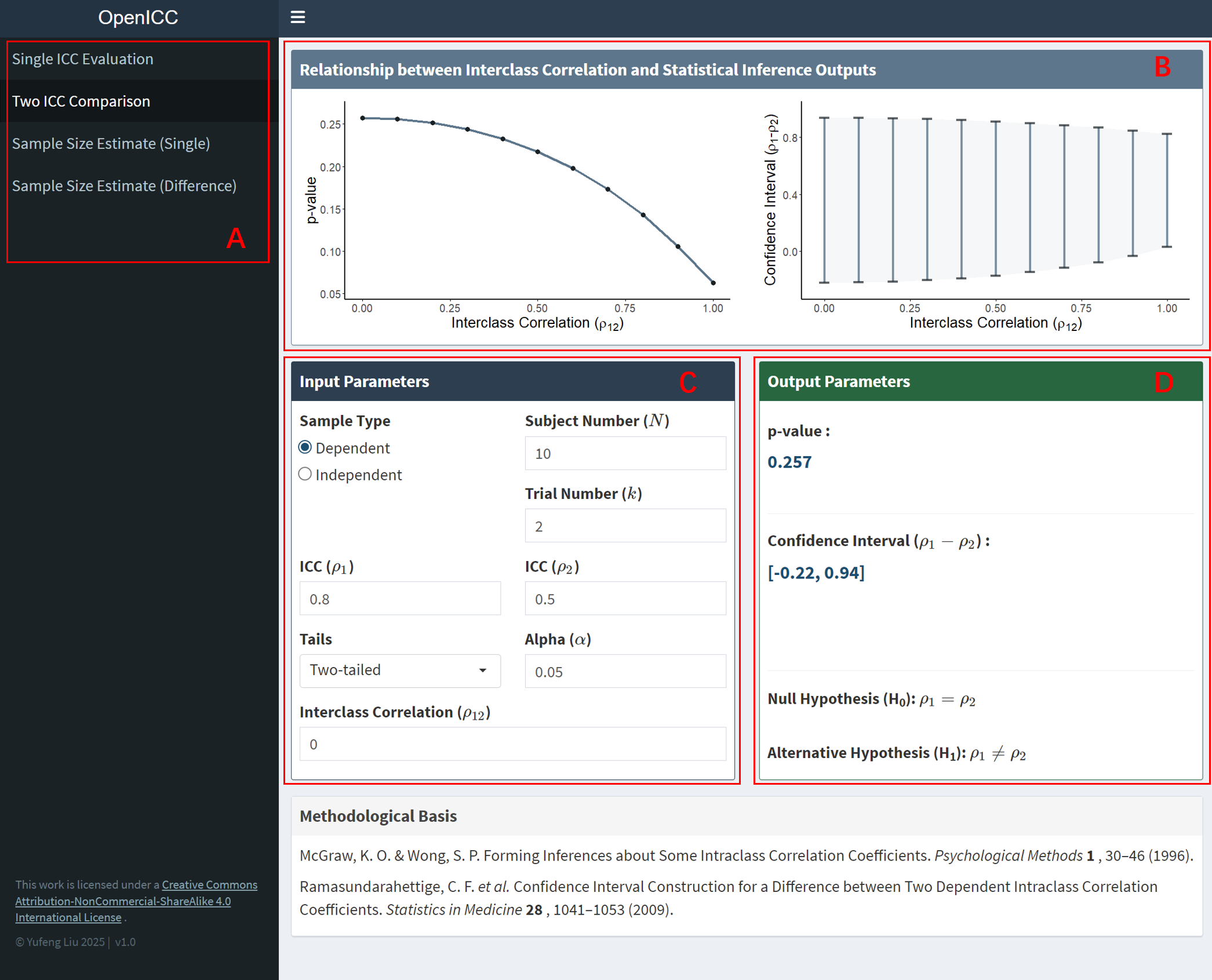}
  \caption{The user interface of the OpenICC application, illustrated using the "Two ICC Comparison" module. The main layout consists of (A) the navigation sidebar, (B) the visualization panel (where applicable), (C) the parameter input panel, and (D) the statistical output panel.}
  \label{app_overview}
\end{figure}

OpenICC is a web-based application that enables ICC inference without requiring any programming expertise. Built on the R Shiny framework, it offers a streamlined, user-friendly interface organized into four modules, each designed for a distinct analytical function (Figure \ref{app_overview}). These modules are described in detail below.  OpenICC is publicly available at   \href{https://nelab.shinyapps.io/OpenICC/}{https://nelab.shinyapps.io/OpenICC/}.

\paragraph{Single ICC Evaluation} module is designed for hypothesis testing of a single ICC. User is allowed to input the sample ICC and specify the number of subjects ($N$), number of trials ($k$), and the null and alternative hypotheses. In addition, users can define the significance level ($\alpha$) and select either a one-tailed or two-tailed test. Based on these parameters, the outputs include the significance level (p-value) of the sample under null hypothesis and the two-sided confidence interval for the population parameter. 

\paragraph{Two ICC Comparison} module evaluates whether two ICCs differ significantly. The interface allows users to specify the number of subjects ($N$), the number of repeated measurements ($k$), and the two ICCs under comparison ($\rho_1$ and $\rho_2$). Users then indicate whether the samples are dependent or independent; for dependent designs, the interclass correlation ($\rho_{12}$) must also be provided. The module supports both one- and two-tailed tests, along with user-defined significance levels ($\alpha$). It returns the p-value and the two-sided confidence interval for the difference ($\rho_1 - \rho_2$). In addition, the module offers a visualization showing how the p-value and confidence interval limits vary as a function of the assumed interclass correlation, enabling users to assess the robustness of their inference.

\paragraph{Sample Size Estimate (Single)} computes the minimum sample size ($N$) required to estimate a single ICC with a prespecified level of precision and statistical power. Leveraging functionality from the R package \verb|ICC.Sample.Size|\cite{rathbone_iccsamplesize_2015}, the module requires users to specify the number of repeated measurements per subject ($k$), the anticipated population ICC ($\rho_1$) under the alternative hypothesis, and the ICC value under the null hypothesis ($\rho_0$). Users also select the test type (one- or two-tailed), the significance level ($\alpha$), and the desired power ($1 - \beta$). The module then returns the minimum sample size ($n$) needed to achieve the specified design parameters.

\paragraph{Sample Size Estimate (Difference)} computes the minimum sample size required to detect a statistically significant difference between two ICCs. The input panel prompts users to specify key design parameters, including the sample type (dependent or independent), the number of repeated measurements ($k$), and the anticipated ICC values ($\rho_1$ and $\rho_2$). For dependent samples, users must also provide the expected interclass correlation ($\rho_{12}$). Statistical criteria are set by choosing a one- or two-tailed test, defining the significance level ($\alpha$), and specifying the desired power ($1 - \beta$). The module then returns the minimum sample size ($N$) needed to reliably detect the ICC difference under the specified design.

\subsection{Verification of Reported ICC Inferences in Test-Retest Studies  }

We used OpenICC to re-evaluate the inferences in the 11 articles (out of 21) that provided sufficient information for verification. Table \ref{summary_verification} summarizes, for each study, the inference type (single ICC evaluation, two ICC comparison or both), the number of recalculated ICC inferences (Recalculated ICC), and the number of consistent ICC inferences between the original statements and the recalculations in each article (Consistent ICC).

We defined the "Consistent Rate" as "Consistent ICC" divided by "Recalculated ICC", reflecting the proportion of recomputed inferences that were compatible with the original statements.

Across the reviewed studies, a total of 283 inferences were recalculated, of which 123 were consistent with the original conclusions, yielding an overall consistency rate of 43.46\%. At the study level, only two of the eleven articles demonstrated fully consistent ICC inferences. In contrast, two articles showed no valid inferences among any of their evaluated statements.

\begin{table}[H]
  \centering 
  \caption{Summary of ICC Verifications in \textit{NeuroImage}}
  \includegraphics[width=\textwidth, height=\textheight, keepaspectratio]{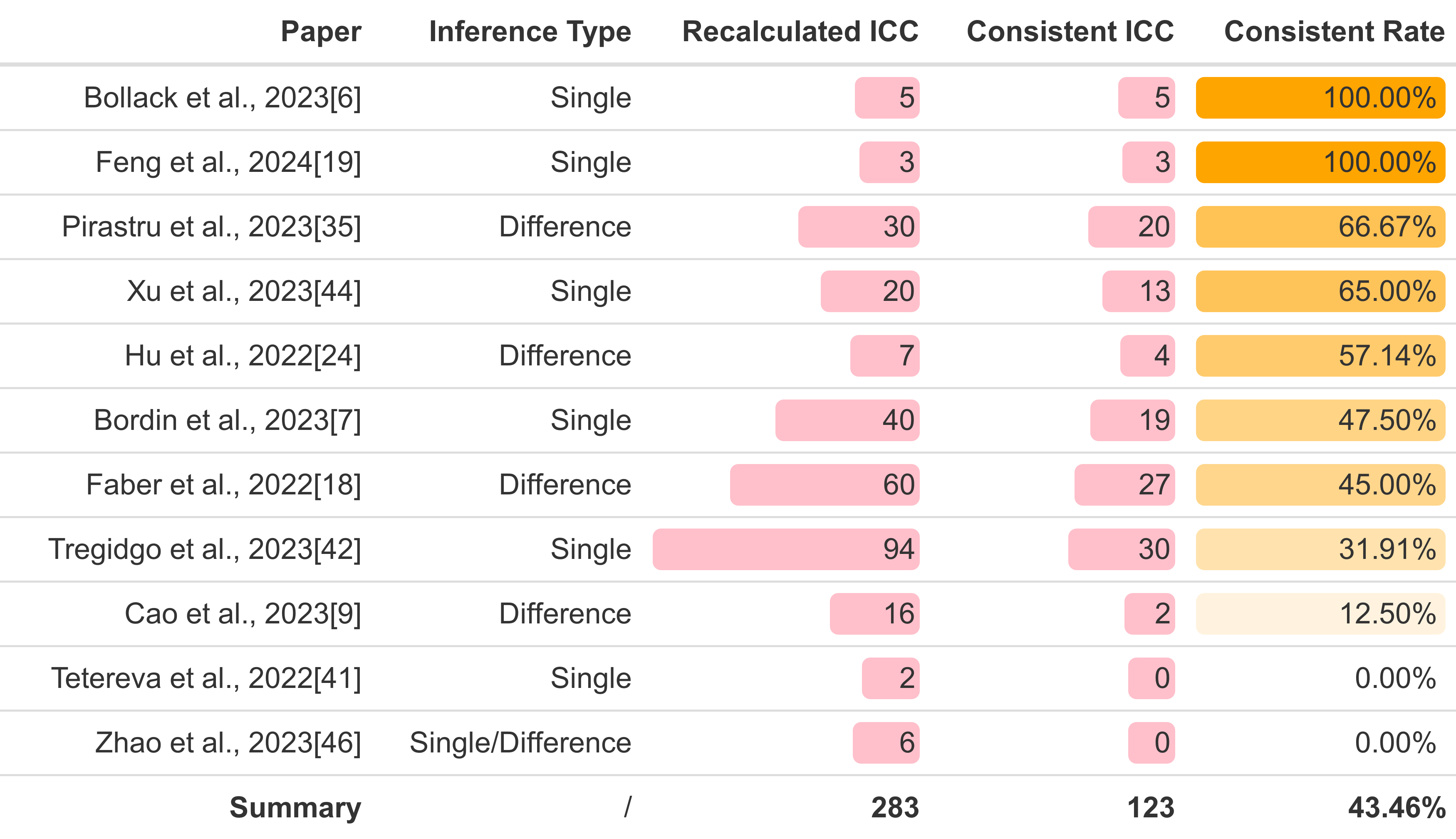}
  \label{summary_verification} 
  \vspace{1ex}
  \parbox{1\textwidth}{
  \small
  \textbf{Notes:} \textbf{Inference Type} specifies the inference category verified in each study. The type include: \textit{Single} = evaluation of a single ICC; \textit{Difference} = comparison of two ICCs; \textit{Single/Difference} = both types included. \textbf{Recalculated ICC} represents the total number of inferences verified in each study. \textbf{Consistent ICC} counts 
the recalculated ICCs that were compatible with the original statements. \textbf{Consistent Rate} is normalized \textbf{Consistent ICC}. The \textbf{Summary} row displays aggregate totals.
  }
 
\end{table}

\section{Discussion}\label{discussion}
\subsection{Violation of the Independence Assumption in Region-wise ICC Comparisons}\label{Misuse_of_Statistical_Tests_on_Dependent_Brain_Region_Data}

A critical methodological concern identified in these studies is the violation of the independence assumption. As noted in Results, several articles conducted between-group tests, though the ICCs were derived from the same dataset \cite{aye_test-retest_2022, kennedy_reliability_2022, wang_mapping_2024, lin_voxel-based_2024}. For example, the authors treated region-wise ICCs as if they were independent samples and stated in \cite{aye_test-retest_2022} that: 
\begin{quote}
    "Next, we calculated the regional ICC using 57 gray matter ROI’s illustrated for $\mathrm{MT_{sat}}$ and $\mathrm{R2^*}$ in Fig. 6. [\dots] We observed a significantly increased ICC of $\mathrm{MT_{sat}}$ in cortical areas compared to subcortical gray matter regions (z(56) = -2.093; $p$ = 0.036; Fig. 6)."
\end{quote}

While this comparison aims to highlight regional differences, the statistical approach is fundamentally flawed. Calculating means and variances across these 57 regions implies that they represent random, independent samples from a population. However, because these measurements originate from the same individuals, they share inherent physiological and anatomical correlations. By treating these correlated regions as independent, the effective degrees of freedom are artificially inflated, leading to an underestimation of the standard error. Consequently, this violation of statistical assumptions  dramatically increases the risk of Type I errors, yielding spurious statistical significance where no meaningful difference exists. 

To statistically compare the distributions of ICC values between two groups of regions (e.g., cortical vs. subcortical regions)  derived from the same cohort, a subject-level bootstrap procedure is recommended. Unlike independent sample t-tests, this non-parametric approach accounts for the dependency of ICC estimates introduced by the shared subject pool. The procedure involves:
\begin{enumerate}
    \item  Resampling $N$ subjects with replacement from the original dataset to generate a bootstrap sample;
    \item Recalculating the ICCs for all cortical and subcortical regions based on this new sample;
    \item Computing the mean difference between the cortical and subcortical ICCs for this iteration;
    \item Repeating this process (e.g., $B=1000$ times) to construct an empirical sampling distribution of the difference.
\end{enumerate}
 Statistical inference is then performed by examining whether the 95\% confidence interval of this difference distribution includes zero. This method rigorously tests whether the reliability in one group of regions is systematically higher than in the other, while preserving the underlying correlation structure of the data.

\subsection{Incorrect Use of Bootstrap Resampling in ICC Comparisons}\label{bootstrap}

As detailed in the Results, one study applied a two-sample t-test to compare two distributions of ICC values generated via bootstrap, treating the number of bootstrap iterations as the sample size \cite{hu_new_2022}. This approach is methodologically flawed in three aspects: 
\begin{enumerate}
\item Dependency: Bootstrap estimates derived from the same dataset are not independent samples;
\item Normality: Sampling distributions of ICCs are typically skewed, violating the normality assumption of the t-test;
\item Statistical power inflation: Most critically, this method treats the number of bootstrap iterations (e.g., $B=1600$) as the sample size ($N$). 
\end{enumerate}

The appropriate bootstrap procedure for comparing two dependent ICCs is to generate the sampling distribution of their difference and base statistical inference on the corresponding confidence interval. Let ($\theta = \rho_{1} - \rho_{2}$) denote the population difference between the two ICCs. From the observed data of ($n$) subjects, we first compute the sample statistics  ($r_{1}$) and ($r_{2}$), yielding the observed difference ($\hat{\theta} = r_{1} - r_{2}$). For each subject ($i$), let ($\mathbf{X}_{i}$) denote the complete vector of repeated measurements, defined as
($\mathbf{X}_{i} = (X_{i1}, X_{i2}, \ldots, X_{ik_{1}},X_{i,k_{1}+1}, X_{i,k_{1}+2}, \ldots, X_{i,k_{1}+k_{2}}$)
where ($k_{1}$) and ($k_{2}$) are the numbers of repeated measurements for the two approaches, respectively.

The correct resampling strategy is therefore to resample subjects as intact units—preserving the joint structure of ($\mathbf{X}_i$)—to construct the empirical distribution of ($\hat{\theta}$). The bootstrap algorithm proceeds as follows:

\begin{enumerate}
\item Resample $n$ subjects with replacement from the original dataset $\{\mathbf{X}_1, \dots, \mathbf{X}_n\}$ to create a bootstrap sample $\{\mathbf{X}_1^*, \dots, \mathbf{X}_n^*\}$.
\item Calculate the ICCs for this bootstrap sample to obtain $r_1^*$ and $r_2^*$.
\item Compute the difference statistic: $\hat{\theta}^* = r_1^* - r_2^*$.
\item Repeat this process $B$ times to generate the bootstrap distribution of the difference: $\hat{\theta}^*_1, \hat{\theta}^*_2, \dots, \hat{\theta}^*_B$.
\end{enumerate}
Statistical inference is then performed by constructing a confidence interval (CI) from this distribution. Using the percentile method, let $\hat{\theta}^*_{q}$ represent the q-th quantile of the bootstrap distribution. The $100(1 - \alpha)\%$ confidence interval for the difference $\theta$ is:
\begin{equation}
\mathrm{CI} = [\hat{\theta}^*_{\alpha/2}, \quad \hat{\theta}^*_{1-\alpha/2}]
\end{equation}
If this interval excludes zero, the difference between the two ICCs is considered statistically significant at the $\alpha$ level.

\subsection{Magnitude-Dependent Uncertainty in ICC Comparisons}

Because the ICC is defined on a bounded scale and estimated from variance components, its sampling distribution is often asymmetric. As a result, uncertainty around an ICC estimate is not uniform across its range: estimates near the upper bound tend to have smaller uncertainty and narrower confidence intervals, whereas mid-range ICC values typically exhibit greater variability. Consequently, point estimates alone fail to convey how uncertainty changes across the ICC scale. This has an important implication: the same absolute difference between two ICC point estimates may correspond to markedly different levels of statistical evidence, depending on their magnitude. 

Using the \textit{Two ICC Comparison} module in OpenICC, consider a study with $N = 28$ and $k = 2$ (assuming $r_{12} = 0$). A difference of 0.10 between two high ICCs (e.g., 0.95 vs. 0.85) is statistically significant and yields a confidence interval for the ICC difference that excludes zero ( $p=0.04$, 95\% CI: $[0.01, 0.25]$). However, the same numerical difference between two moderate ICCs  (e.g., 0.75 vs. 0.65) is not statistically significant and yields a confidence interval that includes zero  ($p=0.47$, 95\% CI: $[-0.18, 0.40]$).   

This further demonstrates that point estimates alone are insufficient to characterize differences at the population level. To determine whether an observed numerical discrepancy reflects a true difference between population ICCs, it is essential to report the confidence interval, or equivalently, to conduct a formal significance test, for the ICC difference.

\subsection{Trade-offs Between Sample Size and Number of Retests}

Many test-retest studies rely on a single retest session. However, increasing the number of retests can substantially reduce the required sample size. When $r_1$ and $r_2$ lie in $[0,1]$ and other parameters are held constant, the inequality constraints in Equation \eqref{eqn}\eqref{eqn:independent_ss} imply that the minimum required sample size decreases monotonically as the number of retests $k$ increases. Thus, for a fixed significance level ($\alpha$) and statistical power, incorporating additional retests can improve design efficiency by reducing the number of participants required. 

To illustrate this effect, we use the\textit{ Sample Size Estimate (Difference)} module of OpenICC, with the following parameters: $r_1= 0.8$ , $r_2 = 0.6$, $r_{12}= 0$ , $\alpha= 0.05$, and power $=0.8$. Under these conditions, a design with two retests ($k=2$) requires 96 participants. Increasing the number of retests to three ($k=3$) reduces the required sample size to 64 participants, and a further increase to four retests ($k=4$) lowers the requirement to 54 participants.

Importantly, the gains from increasing $k$ are subject to diminishing returns. In this example, increasing $k$ from 2 to 3 reduces the required sample size by 32, whereas increasing $k$ from 3 to 4 yields a reduction of only 10 subjects. Moreover, each additional retest imposes a greater burden on participants, potentially elevating the risk of attrition (drop-out). Therefore, when considering additional retests to reduce sample size, researchers should balance the efficiency gain against these practical costs. 

\subsection{Assumption of Equal Retest Numbers and Scope of Applicability}

In OpenICC, we assume an equal number of repeated measurements for the two methods (\(k_1 = k_2\)). Although statistical comparisons of ICCs with unequal numbers of retests are theoretically possible~\cite{donner_testing_2002}, such analyses require estimation of \( \theta = E(Z_{r_1} - Z_{r_2}) \), which in turn necessitates access to raw, subject-level data.

\section{Conclusion}\label{conclusion}

This study systematically examined the use of intraclass correlation coefficients (ICCs) in test–retest reliability analyses in neuroimaging research and identified widespread non-standard practices in articles published in \textit{NeuroImage}. Reanalysis using standardized inferential approaches revealed that many reported ICC-based conclusions may be unreliable when confidence intervals and formal statistical tests are omitted. To address these issues, we provide clear methodological guidelines, standardized inference procedures, and an openly accessible toolbox for ICC-based hypothesis testing and sample size estimation. Although our empirical review focused on \textit{NeuroImage}, the methodological concerns and recommendations outlined here are broadly applicable to test–retest studies across neuroscience and biomedical research. By facilitating transparent, statistically rigorous, and standardized ICC inference, this work aims to improve the reproducibility and credibility of reliability assessments in neuroimaging studies.

\section{Acknowledgements}

This work was supported by the MOST 2030 Brain Project (2022ZD0208500) and the National Natural Science Foundation of China (82472097). X Hong was supported by Science and Technology Commission of Shanghai Municipality (23Y11900500).

\printbibliography
\end{document}